\documentclass[twocolumn,amsmath,amssymb,showkeys]{revtex4-1}
\usepackage{subfigure}
\usepackage{wrapfig}
\usepackage{graphicx}
\usepackage{dcolumn}
\usepackage{bm}
\usepackage{amsmath}
\usepackage{amsfonts}
\usepackage{amssymb}
\usepackage{color}
\usepackage{mathrsfs}
\makeatletter

\begin{document}
\title{Anomalous snapping behavior in asymmetrically constrained elastic strips}
\author{Tomohiko G. Sano$^{1,2}$}
\email[To whom correspondence should be addressed.\\]{tomohiko@gst.ritsumei.ac.jp}
\author{Hirofumi Wada$^{1}$}

\affiliation{$^{1}$Department of Physical Sciences, Ritsumeikan University, Kusatsu, Shiga 525-8577, Japan}
\affiliation{$^{2}$Research Organization of Science and Technology, Ritsumeikan University, Kusatsu, Shiga 525-8577, Japan}


\begin{abstract}
When a flat elastic strip is compressed along its axis, it is bent in one of two possible directions via spontaneous symmetry breaking and forms a cylindrical arc, a phenomenon well known as Euler buckling.
When this cylindrical section is pushed in the other direction, the bending direction can suddenly reverse.
This instability is called snap-through buckling and is one of the elementary shape transitions in a prestressed thin structure. 
Combining experiments and theory, we study snap-buckling of an elastic strip with one end hinged and the other end clamped.
These asymmetric boundary constraints break the intrinsic symmetry of the strip, generating rich exotic mechanical behaviors including
largely hysteretic but reproducible force responses and switch-like discontinuous shape changes. 
We establish the set of exact analytical solutions that fully explain all of our major experimental and numerical findings.
Asymmetric boundary conditions arise naturally in diverse situations when a thin object is in contact with a solid surface at one end, but their profound consequences for the buckling mechanics have been largely overlooked to date.
The idea of introducing asymmetry through boundary conditions would yield new insight into complex and programmable functionalities in material and industrial design.
\end{abstract}
\maketitle

\section{Introduction}
Symmetry is one of the most fundamental concepts in the natural sciences.
Symmetry breaking is often the first step in the development of a variety of spatial and temporal patterns from a featureless background in many natural systems. 
Consideration of the symmetries in a given system is thus essential to classify, for example, the characteristics of phase transition phenomena in various fields ranging from condensed matter physics~\cite{Goldenfeld, Chaikin} and high-energy physics~\cite{Peskin} to living systems~\cite{Naganathan, Ball, Tjhung}. 
To understand emergent complex patterns that are diverse on both length scales and time scales, we usually focus on their {\it intrinsic} symmetries, which can be spontaneously broken in a bulk system. 
Asymmetries, on the other hand, are often introduced into a system in an {\it extrinsic} manner, for example, by external fields and, most importantly here, as boundary conditions (i.e., mechanical or geometric constraints). 
A typical class in which asymmetry imposed at boundaries plays an important role may be the dynamics of a flagellum or cilium of microswimmers
~\cite{Wiggins_1998,Camalet_2000,Yu_2006,Lauga_Powers_2009}. 
One end of the rod-like structure is attached to the body, typically either by a clamped or hinged attachment, whereas the other end is free. 
This boundary condition asymmetry, when driven at one end or internally, can break the inherent symmetry of a system itself and couple it with the surrounding fluid for propulsion. 

\begin{figure}
\centering
\includegraphics[scale = 0.25]{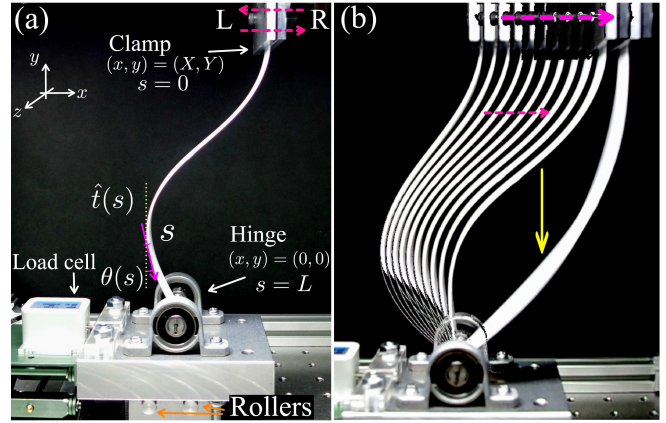}
\caption{Experimental photographs of an asymmetrically constrained elastic strip ($\epsilon_y = 0.16$). (a) Experimental apparatus with the definition of key variables. $\theta(s)$ and $\hat{t}(s)$ are the bending angle from the $y$ axis and the tangent vector at arc length $s$, respectively. 
The strip is initially bent leftward. The clamped end is moved from left (L) to right (R) (forward process) and is moved back (from R to L) (backward process). See also Movie S1.
(b) Stroboscopic pictures of the forward process. Dashed and solid arrows represent the directions of motion and snap-through, respectively.}
\label{setup}
\end{figure}

A thin geometric motif such as a plate or strip is a building block for more complex solid structures in nature
~\cite{darcy_thompson,Harrington_NatCom_2011,ghosal_prl_2012}, 
industry
~\cite{gecko_1,gecko_2,gecko_3,gecko_4,gecko_5,miller_jss_2015},
and everyday life
~\cite{p_reise_pnas_2014,ribe_pre_2003,brun_prl_2015,mahadevan_proc_1996,morigaki_prl_2016}, and is currently a target of active research in various scientific fields
~\cite{Bazant_book,landau_elasticity,audoly_book,Gomez_NatPhys_2016,euler_buckling1,wang_review_1986,he_etal_1997,plaut_2011,powers_rmp_2011,Bosi:2015gx,Bigoni}. 
In classical Euler buckling, one of two possible bending directions is selected by spontaneous symmetry breaking. 
The bending direction can, however, be reversed by indenting the cylindrical section; the reversal occurs as another elastic instability, called snap-through buckling (or snap buckling).
Although snap buckling has a long history of both study and industrial applications, for example, in keyboards and switching devices
~\cite{Goncalves_2003,Daynes_2008,Hung_1999,Plaut_2009,Plaut_2015,Plaut1979,Camescasse:2014hh},
it is currently receiving increasing attention~\cite{Giomi_2012,Bende_2015PNAS,Pandey_2014_EPL} in various scientific fields, for example, in studies of mechanical metamaterials~\cite{Reis_EML_2015,Reis:2015hb,Raney30082016,Marthelot:2017el}, small robots~\cite{Yamada:2010ky}, or nastic motions in plants~\cite{Forterre_Nature_2005,Forterre_JEB_2013,Skotheim_science_2005, Vincent_PRS_2011, Smith_JTB_2011}. 
To highlight the impact of asymmetric boundary constraints on the mechanics of an elastic geometric structure, here we investigate boundary-driven snap-through buckling of an intrinsically flat elastic strip.

In most studies of snapping problems, the strip's ends are symmetrically constrained, typically by either clamping or hinging~\cite{Bazant_book}. 
In this paper, we provide simple snap-buckling behavior induced solely by the asymmetrically controlled boundaries. 
More specifically, we study the snap mechanics of an elastica with one end hinged and the other end clamped by shearing them apart [Fig.~\ref{setup}(a)]. 
Our controlled physical experiment reveals that the asymmetric boundaries produce highly anomalous, largely hysteretic but reproducible force responses.
We establish a set of exact analytical solutions for elastica under the clamped--hinged constraints and for an arbitrary horizontal strain $\epsilon_x$, which, to our knowledge, has never been reported. 
These analytical solutions, as well as corroborative numerical simulations, fully explain all the major findings in our experiments.
Furthermore, we also construct the exact bending energy landscape of an inextensible elastica, from which we can deduce a scaling law in the fast dynamics during snapping.
This result may complement a recent study of snap-buckling dynamics with variable clamped--clamped boundary conditions~\cite{Gomez_NatPhys_2016}.
Importantly, we show that all the snapping properties newly reported here are absent in the standard setup of symmetric boundary conditions.
Asymmetric boundary conditions arise naturally in diverse situations when a thin object is in contact with a solid surface at one end~\cite{gecko_1,gecko_2,gecko_3,gecko_4,gecko_5,miller_jss_2015,Mogilner_1996,massa_gilroy_2003,Daniels_2006,persson_book,popov_book,sano_yamaguchi_wada_letter,mahadevan_friction_2015,klein_nature_1994}, but their profound consequences for the buckling mechanics seem to be largely overlooked to date.

\begin{figure}[t]
\centering
\includegraphics[scale = 1.33]{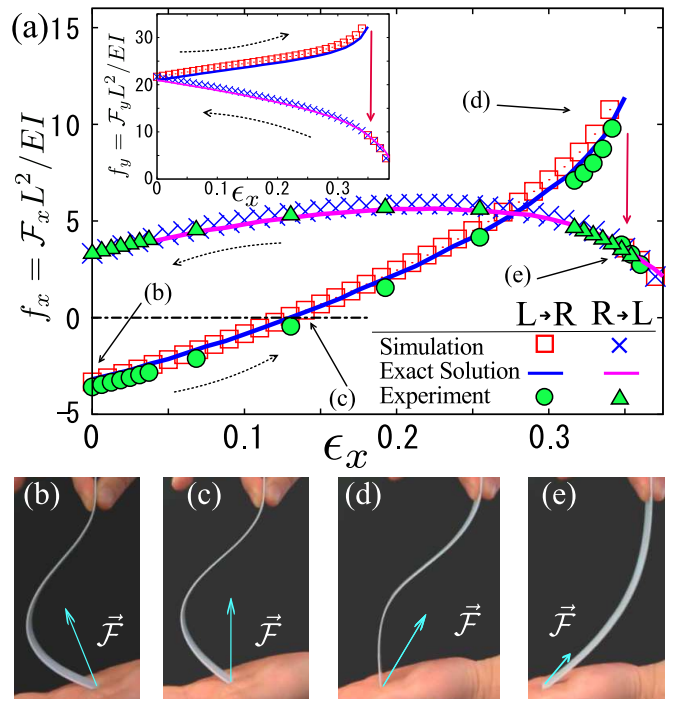}
\caption{Force vs. strain relations of an asymmetrically constrained elastic strip and its corresponding shapes. 
(a) Force vs. strain curves obtained from experiments, exact solutions, and numerical simulations for $\epsilon_y\simeq 0.1$. 
Main panel shows the rescaled horizontal force $f_x = {\mathcal F}_x L^2/EI$; inset shows the rescaled vertical force $f_y = {\mathcal F}_y L^2/EI$. 
Dashed arrow denotes the direction of sliding motion. 
Filled circles and triangles are the experimental results in the forward (L to R) and backward (R to L) processes, respectively.
Open squares and $\times$ symbols are data from our numerical simulations in the forward (L to R) and backward (R to L) protocols, respectively. 
Solid lines are the predictions from the exact analytical solutions.
(b)--(e) Illustrative snapshots that show typical configurations of an elastic strip during the processes described in (a).
In this experiment, the top end of a strip is picked up with the fingers while its bottom end is in contact with the palm of the hand.
Here, the combination of the fingers and palm mimic clamped--hinged boundary conditions.
Note that (b)--(e) are for illustration purpose only, and the data shown in (a) are actually obtained in the controlled setup described in Fig.~\ref{setup}.}
\label{fig:force_strain}
\end{figure}

\section{Experiment}
\subsection{Experimental apparatus}
We cut a rectangular strip from a plastic sheet made of rigid polyvinyl chloride (Acrysunday, No. 500).
Our strip is intrinsically flat; it has a uniform thickness $h = $1[mm] and width $w = 20$[mm], and three total lengths, $L = 156, 161, 172$ [mm] (this is the effective length when it is mounted in the experimental system).
The strip is set vertically; its top end is clamped, and its bottom end is hinged [see Fig.~\ref{setup}(b)].
Throughout our experiment, the origin of the coordinate system is set at the hinged end, and the vertical height of the clamped end is fixed at $Y=145$ [mm].
The initial vertical strains $\epsilon_y \equiv 1 - Y/L$ thus vary for strips of different lengths as $\epsilon_y = 0.071, 0.099$, and 0.16 for strips with $L = 156, 161$ and 172 [mm], respectively,.
The horizontal position of the clamped top, $(X, Y)$, is controlled by a stepping motor (Oriental Motors, ARM46AC); $X$ is changed sufficiently slowly to maintain mechanical equilibrium at each position [Fig.~\ref{setup}(b)]. 
The Young's modulus $E$ of our strips is determined by measuring the critical force for ${\mathcal F}_x$ at $X = 0$, which yields $E = 3.7, 3.3,$ and $3.9$ [GPa] for strips with $L = 156, 161$, and 172 [mm], respectively; these are typical values for rigid polyvinyl chlorides. 
All the measured force data are shown in units of $EI/L^2$, and the moment of inertia of a strip is $I = h^3w/12$. 
At the hinged boundary, the strip's end is glued to an empty hexagonal shaft (TAMIYA, 2 [mm] diameter, 72 [mm] length), which rotates around two bearings (TAMIYA, inner diameter 2 [mm], outer diameter 6 [mm]). 
The bearings are inserted into two bearing supports (Uxcell Japan, A14071400UX0285, inner radius 8 [mm], outer radius 12 [mm]), which are fixed on a horizontally movable hinge stage. 
We confirm that the shaft placed inside the bearing rotates almost freely; thus, the rotational friction forces can safely be neglected for our purpose. 
The hinge stage is placed on two slender cylinders and tightly connected to the load cell (Kyowa Dengyo, LTS-2KA) to measure the horizontal force acting on the strips [Fig.~\ref{setup}(a)]. 
The apparatus is sufficiently rigid; the horizontality of the load cell, hinge stage, and stepping motor is ensured during snapping and throughout the entire measurement.

We set the initial position of the clamped end as $X < 0$ to bend the strip leftward [Fig.~\ref{setup}(b)].
We then move (slide) the clamped end by increasing $X$ by $1-10$ [mm] per step [the motion is thus from left (L) to right (R)], maintaining the mechanical equilibrium of the system at each step (the forward process).
At the critical horizontal strain $\epsilon_x^*$, the strip snaps, and the buckling direction reverses. 
We then switch the direction of motion of the clamped end, returning it to the initial position at the same speed (the backward process).
See {Appendix}~\ref{sec:exp_app}, for experimental details.

\subsection{Experimental results}
In the main panel of Fig.~\ref{fig:force_strain}(a), we plot the rescaled horizontal force $f_x= {\mathcal F}_xL^2/{EI}$ as a function of the horizontal strain $\epsilon_x \equiv X/L$ for $\epsilon_y = 0.099$, where filled circles and triangles denote data obtained during the forward and backward processes, respectively. 
When the strip is at rest at $\epsilon_x = 0$ [Fig.~\ref{fig:force_strain}(b)], it is pushed leftward from the hinge; i.e., $f_x < 0$. 
As the clamped end moves (as $X$ increases), the magnitude of $f_x$ decreases and approaches zero, and across $\epsilon_x\simeq 0.13$ its sign changes; i.e., the loading changes from compressive to tensile [Fig.~\ref{fig:force_strain}(c)]. 
Note, however, that no shape instability is observed at this point. 
The horizontal force $f_x$ increases continuously as $\epsilon_x$ becomes larger beyond the point at which $f_x=0$, and the strip eventually snaps when the hinged end becomes perpendicular to the flat bottom stage at the critical strain $\epsilon_x ^*= 0.34$ [see Fig.~\ref{fig:force_strain}(d)(e)]. 
At this transition point, the horizontal force from the substrate $f_x$ decreases discontinuously, whereas its direction is unchanged. 
After the snap, the direction of the sliding motion of the clamped end is reversed. 
The force curve in this backward process differs considerably from that observed in the forward process;
the force response in this cyclic process is largely hysteretic, but at the same time it is absolutely reproducible, as is typical of athermal bistable systems.
Interestingly, although the configurations of the two ends are exactly the same, the configuration of the buckled strip is now a mirror image of the original (the direction of buckling is opposite). 
This is reflected in the main panel of Fig.~\ref{fig:force_strain}(a); at $X=0$, $f_x$ has two values of the same magnitude but with opposite signs.

\section{Analytical solutions of asymmetrically constrained strip}
The force response observed in our experiments is anomalous and highly nonlinear.
To rationalize it, we now construct the set of exact solutions of the clamped--hinged elastica for arbitrary $\epsilon_x$. 
Despite the long history of the mathematics of elastica~\cite{euler_buckling1,mitaka_acta_mechanica_2007}, these solutions have, to our knowledge, never been derived. 
Let $\vec{F}(s)$ and the moment $\vec{M}(s)$ be the internal force and moment, respectively, over the cross section of a strip at position $s$, which are exerted by the section of the strip with an arc length greater than $s$ on the section of the strip with an arc length less than $s$. 
Note that $s$ represents the arc length measured from the clamped end. In the absence of any external body forces and moments, the balance of these internal forces and moments leads to the Kirchhoff rod equations~\cite{powers_rmp_2011,audoly_book},
$\vec{F}'(s) = 0$ and $\vec{M}'(s) + \hat{t}(s)\times\vec{F}(s) = 0$,
with the linear constitutive relation $\vec{M}(s) = EI\theta'(s)\hat{z}$ and the tangent vector $\hat{t}(s) = (\sin\theta(s),-\cos\theta(s))$, where the prime ($'$) represents the derivative in terms of $s$, and $\theta(s)$ is the bending angle at $s$ [Fig.~\ref{setup}(a)].
We apply both tangential and horizontal forces at the clamped end: $\vec{F}(0) = \vec{\mathcal F}$. By solving the first equation as $\vec{F}(s) = ({\mathcal F}_x,\mathcal{F}_y)$ and substituting the result into the second equation, we obtain the equation for $\vartheta(\tau)\equiv\theta(L\tau)$ as
\begin{eqnarray}
\ddot{\vartheta}(\tau) = -f_x\cos\vartheta(\tau)-f_y\sin\vartheta(\tau)\label{eq:ddtheta},
\end{eqnarray}
together with the boundary conditions at the top, $\vartheta(0) = 0$, and at the bottom, $\dot{\vartheta}(1) = 0$, where the dot ($\dot{}$) represents the derivative with respect to $\tau \equiv s/L$. 
${f}_x$ and ${f}_y$ are determined from the constraints on the position of the clamped end, $(x(0),y(0)) = (X, Y)$, i.e.,
\begin{eqnarray}
\int_0 ^1 \sin\vartheta(\tau)d\tau = -\epsilon_x,\label{x1_0}~~
1 - \int_0 ^1 \cos\theta(\tau)d\tau = \epsilon_y.
\label{eq:constraints}
\end{eqnarray}
Equations~(\ref{eq:ddtheta}) and (\ref{eq:constraints}) can be solved using elliptic integrals~\cite{mitaka_acta_mechanica_2007,Abramowitz}. 

The exact solutions of Eqs.~(\ref{eq:ddtheta}) and (\ref{x1_0}) are classified into two cases. 
The first case corresponds to a pre-snap shape satisfying $|\epsilon_x| \leq \epsilon_x ^*$. 
Mathematically, this configuration has an inflection point at $s^*$ in terms of the bending angle $\theta(s)$. 
The existence of the inflection point allows two possible, but not symmetric, buckling directions. 
The inflection point vanishes and the two solutions merge at $\epsilon_x=\epsilon_x ^*$, and, as soon as $|\epsilon_x|$ exceeds $\epsilon_x ^*$, the solution is unique and describes a post-snap shape. 
On the basis of this classification, in either case, we are left with two nonlinear algebraic equations consisting of elliptic integrals for two undetermined coefficients $f_x$ and $f_y$, which are readily solved numerically. 
In Fig.~\ref{fig:force_strain}(a), we show our exact analytical solution (solid line), which accurately predicts the experimental result. 
We also performed numerical simulations using a discretized analog of a continuous elastic strip~\cite{discrete_model},
which are shown in Fig.~\ref{fig:force_strain}(a) as squares (forward) and $\times$ symbols (backward). 
(See {Appendices}~\ref{sec:theory_app} and \ref{sec:simulation}, for detailed derivations and details of our numerical method, respectively.)
The numerical simulation data confirm our analytical prediction and agree quite well with our experimental measurements.
Note that in our analytical and numerical analysis, all the relevant parameters are taken from those in our experiments, and 
agreement between them is obtained without any adjustable parameters. 

\begin{figure}[t]
\centering
\includegraphics[scale = 0.95]{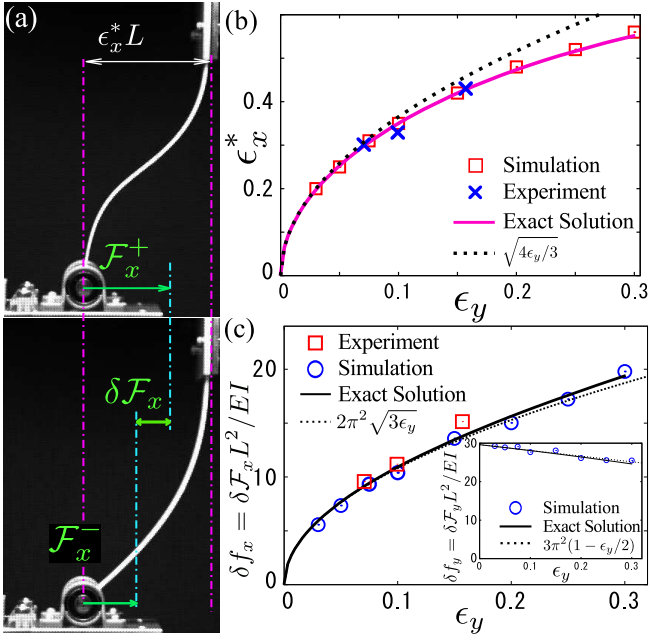}
\caption{Critical horizontal strain and magnitude of force change upon snapping. 
(a) Experimental strip configurations immediately before (top) and immediately after (bottom) the transition, captured by high-speed camera. 
(b) Critical horizontal strain $\epsilon_x ^*$ as a function of preset vertical strain $\epsilon_y$. 
Squares and $\times$ symbols represent data from our numerical simulations and physical experiments, respectively.
Solid and dashed lines are the exact analytical solution and its approximate expression given in Eq.~(\ref{eq:approx_critical_main}), respectively. 
(c) Magnitude of the discontinuous increase in force at the snapping transition: $\delta f_x = \delta {\mathcal F}_x L^2/EI$ (main panel) and $\delta f_y = \delta {\mathcal F}_y L^2/EI$ (inset). 
Circles and squares represent the numerical and experimental results, respectively. 
Solid and dashed lines are the predictions from the exact analytical solution and its approximation given in Eq.~(\ref{eq:force_change_approx_main}), respectively.}
\label{fig:critical_strain}
\end{figure}

To obtain the critical strains for the snapping transition, we now focus on the small strain regime, $\epsilon_y \ll 1$.
The snapping transition is purely geometrically controlled; it occurs when the hinged end becomes vertical. 
Thus, the critical strain for snapping, $\epsilon_x ^*$, can be derived from the condition $\theta(L) = 0$. 
By introducing the angle variable $\varphi$ as $\tan\varphi \equiv f_x/f_y$, Eq.~(\ref{x1_0}) is simplified in the limit $\epsilon_y\ll 1$ as
\begin{eqnarray}
\epsilon_x ^* \simeq \varphi,~~ \epsilon_y \simeq\frac{3\varphi^2}{4},
\label{eq:approx_critical_main}
\end{eqnarray}
which leads to $\epsilon_x ^* \simeq \sqrt{4\epsilon_y/3}$. We plot this in Fig.~\ref{fig:critical_strain}(b), together with the exact result (see Appendix~\ref{sec:theory_app}, for the derivation), which agrees quite well with both the simulation and experimental results.

\subsection{Rapid force change at the onset of snapping}

In the snapping transition, the boundary force necessary to hold the elastic structure changes abruptly from $({\mathcal F}_x ^+, {\mathcal F}_y ^+)$ to $({\mathcal F}_x ^-, {\mathcal F}_y ^-)$. 
In fact, this discontinuous force change may be ubiquitous in solid systems, including those in a wide range of engineering applications.
To know the magnitude of the force changes $\delta{\mathcal F}_x = {\mathcal F}_x ^+ - {\mathcal F}_x ^-$ and $\delta{\mathcal F}_y = {\mathcal F}_y ^+ - {\mathcal F}_y ^-$ in advance could therefore be helpful for assessing the potential risk in different practical situations. 
The discontinuous changes in the external forces are accurately predicted on the basis of our analytical results. 
Taking the difference between the forces of the pre- and post-snapped shapes at the onset of the transition, $\epsilon_x ^*$, we obtain for $\epsilon_y \ll 1$ 
\begin{eqnarray}
\delta f_x \simeq 2\pi^2 \sqrt{3\epsilon_y},~~
\delta f_y \simeq 3\pi^2 \left(1 - \frac{\epsilon_y}{2}\right).
\label{eq:force_change_approx_main}
\end{eqnarray}
These analytical predictions agree quite well with the experimental and numerical results for small $\epsilon_y$ in Fig.~\ref{fig:critical_strain}(c). 
Note that the horizontal force change $\delta f_x$ depends nonlinearly on $\epsilon_y$ as $\sqrt{\epsilon_y}$, suggesting that it is particularly large compared to the linear response for $\epsilon_y \ll 1$.
This rapid force change with controlled nonlinearity could be used in the design of future mechanical systems.

\begin{figure*}[!t]
\centering
\includegraphics[scale = 1.4]{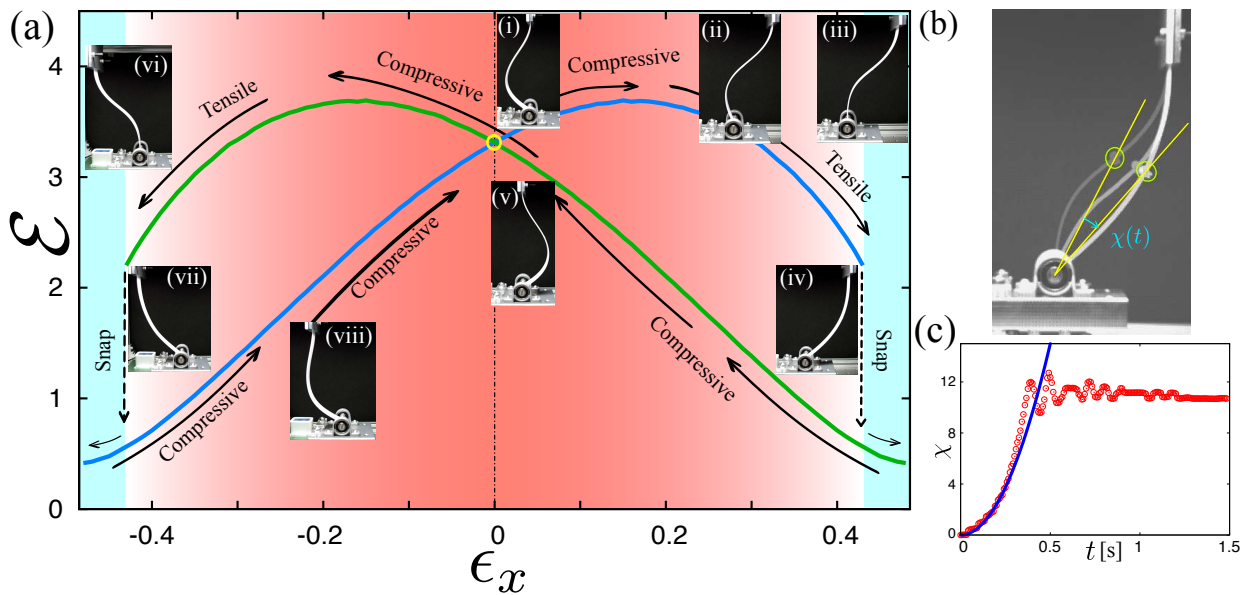}
\caption{Elastic energy landscape and fast snapping dynamics. 
(a) Calculated bending elastic energy as a function of the horizontal strain $\epsilon_x$ for a fixed vertical strain $\epsilon_y = 0.16$ (solid lines). 
Black arrows represent the sliding direction of the clamped end. 
Typical experimental configurations of a strip for different $\epsilon_x$ are shown in insets (i)--(viii). 
Red and blue regions represent the bistable and monostable regions, respectively. 
(b) Stroboscopic picture of the strip during the snap transition and definition of the bending angle $\chi(t)$. (c) Experimental results of blow-up dynamics during the snapping shape transition.
Time evolution of the angle $\chi(t)$ is shown. 
Dashed line is the best fit to the experimental data with our scaling prediction, $\chi\propto t^2$. 
}
\label{fig:highspeed}
\end{figure*}

\section{Elastic energy landscape and snap dynamics}

Using the analytical results, we now reconstruct the landscape of the elastic deformation energy as a function of $\epsilon_x$.
Note that for an inextensible strip, the total elastic energy is the bending energy ${\mathcal E} = \int_0 ^1 \dot{\vartheta}^2(\tau)d\tau/2$ (given in rescaled units). 
In Fig.~\ref{fig:highspeed}(a), we plot the shape of ${\mathcal E}$ for $\epsilon_y = 0.16$, together with the corresponding shapes observed in the experiments [see insets (i)--(viii)]. 
The initial Euler-buckled shapes, bent left- and rightward for $\epsilon_x = 0$, are energetically degenerate [Fig.~\ref{fig:highspeed}(a)(i) and (v)]. 
When, for example, a leftward-bent one is slid toward the right, i.e., $\epsilon_x > 0$, the bending energy initially increases as the strip is subjected to compressional stress.
At approximately $\epsilon_x\simeq 0.2$, the energy reaches a maximum, and the strip is then under tensile loading for larger $\epsilon_x$ [Fig.~\ref{fig:highspeed}(a)(ii)]. 
At approximately $\epsilon_x\simeq0.43$, snap-buckling occurs [Fig.~\ref{fig:highspeed}(a)(iii) and (iv)]; mathematically, the analytical solution with an inflection point no longer exists at this point. 
The energy landscape in the backward process is distinctly different from that in the forward process; the difference is the direct origin of the hysteretic force response that we observed above.
Because of the way our protocol is constructed, the strip's configuration at $\epsilon_x=0$ is now a mirror image of the original [Fig.~\ref{fig:highspeed}(a)(v)].
To restore it to the original configuration, the cycle needs to be repeated for $\epsilon_x < 0$, in which all the configurations are simply mirror images of those observed for $\epsilon_x>0$ [Fig.~\ref{fig:highspeed}(a)(vi), (vii), and (viii)].

From the energy diagram in Fig.~\ref{fig:highspeed}(a), we can infer some dynamical properties at the onset of snapping. 
In Fig.~\ref{fig:highspeed}(a), the red- and blue-shaded areas represent the bi- and monostable regions, respectively. 
At $|\epsilon_x| = \epsilon_x ^*\simeq 0.43$, the upper branch disappears, and snapping occurs. 
This sudden disappearance of the energy branch implies a saddle-node/fold bifurcation~\cite{Pippard}. 
If we add an infinitesimally small strain $\delta\epsilon_x$ to the strip's critical state with $0<\delta\epsilon_x \ll 1$ and $\delta\epsilon_x\equiv\epsilon_x -\epsilon_x ^*$, the state of the strip immediately jumps to the lower branch. 
This fast dynamics can be qualitatively understood according to the following scaling argument. 
Suppose the bending angle $\chi(t)$ is defined precisely, as shown in Fig.~\ref{fig:highspeed}(b). 
The dynamics of $\chi(t)$ could in principle be derived from Newton's equation of motion, which describes the balance of the inertial force and internal elastic force on the strip.
At this scaling level, we assume that it is governed by the second-order nonlinear ordinary differential equation $d^2\chi/dt^2 = g(\chi, \delta\epsilon_x)$ with initial conditions $\chi(0) = 0$ and $d\chi/dt|_{t = 0} = 0$, where $g = g(\chi, \delta\epsilon_x)$ is a nonlinear function of $\chi$ and $\delta\epsilon_x$. Expanding $g$ in terms of $\chi \ll 1$ and $\delta\epsilon_x\ll1$ and leaving only the leading order terms, one finds $g(\chi, \delta\epsilon_x) \simeq a\delta\epsilon_x + b\chi^2$, with constants $a = \partial g/\partial\delta\epsilon_x|_{(0,0)}$ and $b = \frac{1}{2}\partial^2 g/\partial\chi^2|_{(0,0)}$. 
Here we used the general properties of the fixed point, $g(0, \delta\epsilon_x) = 0$ and $\partial g/\partial \chi|_{(0,\delta\epsilon_x)} = 0$~\cite{Strogatz}. 
Thus, the reduced dynamics for $\chi(t)$ is written as $d^2\chi/dt^2 = a\delta\epsilon_x + b\chi^2$, which is the normal form of saddle-node bifurcation. A more sophisticated argument on the normal form in snap-buckling is found in a recent paper by Gomez, {\it et al.}~\cite{Gomez_NatPhys_2016}. From the reduced equation, $\chi$ is predicted to grow initially as $\chi \propto t^2$. Indeed, this blow-up dynamics is confirmed in our experiments using a high-speed camera (Detect, HAS-D71, 2000 fps, Movie S2). In Fig.~\ref{fig:highspeed}(c), we plot the fitting result for $\chi \propto t^2$ as a solid line. Note that the strip exhibits elastic oscillation after the blowup.

\begin{figure}[!b]
\centering
\includegraphics[scale = 1.17]{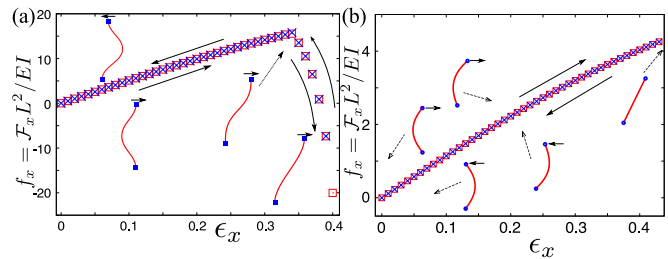}
\caption{
Force vs.~strain relations of a strip with (a) clamped--clamped and (b) hinged--hinged boundary conditions together with typical configurations for different strains for $\epsilon_y= 0.10$. 
Squares and $\times$ symbols represent data obtained from our numerical simulations for the forward (L to R) and backward (R to L) processes, respectively. }
\label{fig:symmetric}
\end{figure}

\section{Discussion: impact of asymmetry}

To demonstrate that our findings are unique to asymmetrically constrained systems, we now compare them to the results obtained from strips with either clamped--clamped or hinged--hinged boundary conditions under the same cyclic protocol for $\epsilon_y = 0.10$. 
In Figs.~\ref{fig:symmetric}(a) and (b), we plot the simulation results for $f_x$ under clamped--clamped and hinged--hinged conditions, respectively, with the corresponding strip's shapes. 
Squares and $\times$ symbols are data for the forward and backward processes, respectively. Although the force $f_x$ for the clamped--clamped case changes sharply at $\epsilon_x\simeq 0.34$, it is continuous, in contrast to the asymmetric case. The force $f_x$ for the hinged--hinged case also changes continuously until the strip is fully stretched at $\epsilon_x = \sqrt{1-(1-\epsilon_y)^2}\simeq0.44$. 
Furthermore, the force responses in the reverse processes are found to be almost identical to those in the forward processes; the mechanical responses are not hysteretic. 
Most importantly, in either case, there are no abrupt force changes, which is a fundamental difference from the asymmetric case.

In studies of snapping, a cylindrical section is typically subjected to a point load at its center. 
It appears that our system could be viewed as one of the two sides of such standard systems. 
However, this is not true for two reasons. 
First, in a point-loading experiment, an asymmetric shape can develop before snap-buckling~\cite{Pandey_2014_EPL,Plaut_2015}. 
Second, the point loading is moment-free; it is thus closer to the hinged condition. 
In principle, point loading with a clamped constraint (and controlled displacement) is possible, but such a condition would be rather artificial compared to our setup. 
Interestingly, in the point-loading experiment (with controlled displacement), a force response similar to that in Fig.~\ref{fig:force_strain}(a) has been reported for off-center indentation, whereas indentation at the center produces a continuous force curve similar to that in Fig.~\ref{fig:symmetric}(b)~\cite{Camescasse:2014hh}. 
Taken together, our present study, as well as those previous attempts~\cite{Camescasse:2014hh, Gomez_NatPhys_2016}, now prove that any asymmetric constraints, either in the boundaries or the indentation point, are the central mechanism for creating a discontinuous response in snapping. 
In other words, the geometric symmetry controls the type of bifurcation, largely independent of the material properties and size. 
This general principle may be useful for distinguishing between a smooth shape change with a continuous force response and a rapid shape change with a discontinuous force response in slender materials.

\section{Summary}\label{sec:summary}

By combining experiment and theory, we studied boundary-induced snapping of an elastic strip with one end clamped and the other end hinged. 
In this setup, snapping occurs when the hinged end becomes perpendicular to the flat bottom stage, and the inflection point of the elastica vanishes. 
Reflecting the asymmetry of the boundary conditions, the force--strain relation exhibits several remarkable features. 
The first is the change in the direction of the horizontal force $f_x$ before the snap instability. 
This occurs because the strip is initially compressed and is then pulled before snapping. 
The second is the discontinuous change in the forces at the onset of the snap transition.
The third is the largely hysteretic but accurately reproducible force vs. strain properties. 
On the basis of the exact solutions that we established, we uncovered the elastic energy landscape over the entire deformation process, including snapping, and extracted the dynamical properties of our peculiar snap-buckling system. 
It should be stressed that the rich exotic mechanical behavior presented in this paper originates purely from the asymmetry in the boundary constraints, which could be explored in other low-dimensional elastic systems.

In closing, we address several possible future directions of our study. 
Recently, new mechanical systems with emergent functionalities, so-called mechanical metamaterials, have been inspired by topological insulators, in which only the surface boundaries can behave as conductors~\cite{huber_2016}. 
Small mechanically asymmetric building blocks are combined, and it is now possible to create materials with unusual order and machine-like functionalities (e.g., using 3D printers)~\cite{coulais_2016, coulais_2017}. 
Several of the properties that we report in this study, such as the switch between compressive and tensile loading, the discontinuous force change upon snap-buckling, and hysteretic behavior, are all uniquely determined and predicted by the externally controllable strain $\epsilon_y$.
Therefore, it would be interesting to adopt asymmetrically constrained but inherently symmetric rods or strips as components of metamaterials. 
By combining these asymmetrically constrained parts to form a large bulk material and scaling up the force--strain relation, it might be possible to create a new mechanical design controlling the compressive and tensile response. 

\begin{acknowledgments}
{We acknowledge financial support in the form of Grants-in-Aid for Japan Society for the Promotion of Science (JSPS) Fellows (Grant No.~28$\cdot$5315) and JSPS KAKENHI (No.~15H03712, No.~16H00815, and "Synergy of Fluctuation and Structure: Quest for Universal Laws in Non-Equilibrium Systems"). We acknowledge T. Yamaguchi and Y. Tanaka for valuable discussions on the experiments. We also thank K. Nakamura and T. Yoneda for technical support. 
}
\end{acknowledgments}

\newpage
\appendix
\section{Experimental details}\label{sec:exp_app}
In this section, we explain the details of the experiments. In Sec.~\ref{sec:exp_force}, we show the detailed procedure for obtaining the force--strain curves. In Secs.~\ref{sec:exp_dfx} and \ref{sec:exp_highspeed}, the procedures for observing $\delta f_x$ and the snap-buckling dynamics are explained, respectively.

\subsection{Procedure for experimental measurement of $\epsilon_x ^*$ and force--strain curves}\label{sec:exp_force}
Before starting the force measurements, we shift the clamped end continuously by forward (L-to-R) and backward (R-to-L) protocols to confirm the left--right symmetry of the system and then return the clamped end to its initial position. We start measuring the force by moving the clamped end a distance of 1 [mm] at minimum and 10 [mm] at maximum. When the clamped end is close to the transition point or close to $X = 0$ (to verify the zero point of $X$), we move the clamped end by 1 [mm]; otherwise, we move it by 10 [mm]. We observe the snap for both positive and negative $X$. Let $X_+ > 0$ and $X_- < 0$ be the positions of the snap in the forward (L-to-R) and backward (R-to-L) processes, respectively. We define the zero of $X$ as $X_0 \equiv (X_+ + X_-)/2$, which is 1 [mm] at most.
By deducting the offset $X_0$ from the observed $X_+$, $\epsilon_x ^*$ is determined as $\epsilon_x ^* = (X_+ - X_0)/L$. Note that the moving speed is about 1 [mm/s], and we wait more than 10 s for relaxation after every moving step so that the system reaches mechanical equilibrium. The force data are recorded by a data logger (Kyowa Dengyo, EDX10-B, EDX14-A). 

\subsection{Procedure for experimental measurement of $\delta f_x$}\label{sec:exp_dfx}
We measure $\delta f_x$ after recording the force--strain curve for each strip. We first move the clamped end to $X \lesssim X_+$ to realize the critical state before snapping. We then start recording the force using the load cell at 100 [Hz], shifting the clamped end by 1 [mm]. After snap-buckling occurs and the strip becomes stationary, we stop recording the force. The force-released onset of the snap is defined as the difference between the maximum and minimum values in the recorded time series data of the force. To obtain $\delta f_x$ for each experiment, we rescale the data by $EI/L^2$, which is used to obtain the force--strain curve for each strip.

\subsection{Procedure for observing snap-buckling dynamics}\label{sec:exp_highspeed}
Using a high-speed camera, we record a movie showing the onset of snap-buckling. We attach a plastic tip to the midpoint of the strip with the horizontal prestrain $\epsilon_y = 0.16$. We move the clamped stage close to the critical strain as $X \lesssim X_+$. Then, shifting the clamped end by 1 [mm], we record the dynamics using the high-speed camera at 2000 frames per second.

\section{Theoretical basis}\label{sec:theory_app}
We outline the derivation of our exact solution presented in the main text. 
Our starting point is 
\begin{eqnarray}
\ddot{\vartheta}(\tau) = -f_x\cos\vartheta(\tau)-f_y\sin\vartheta(\tau)\label{eq:ddtheta_app},
\end{eqnarray}
together with the boundary conditions at the top, $\vartheta(0) = 0$, and at the bottom, $\dot{\vartheta}(1) = 0$, and the constraints
\begin{eqnarray}
\int_0 ^1 \sin\vartheta(\tau)d\tau &=& -\epsilon_x,\label{x1_0_app}~
1 - \int_0 ^1 \cos\theta(\tau)d\tau = \epsilon_y.
\end{eqnarray}
Let us rewrite Eq.~(\ref{eq:ddtheta_app}) as
\begin{eqnarray}
\ddot{\phi}(\tau) &=& - f \sin(\phi(\tau)),\label{eq_theta_0}\\
f&\equiv&\sqrt{f_x ^2 + f_y ^2},\\
\tan\varphi&\equiv&f_x/f_y,
\end{eqnarray}
together with the boundary conditions $\phi(0) = \varphi$ and $\dot{\phi}(1) = 0$, where
we introduce the new angular variable $\phi(\tau) \equiv \vartheta(\tau) + \varphi$. After multiplying Eq.~(\ref{eq_theta_0}) by $\dot{\phi}$, we integrate Eq.~(\ref{eq_theta_0}) from $\tau$ to $1$ as
\begin{eqnarray}
-\frac{1}{2}\left\{\dot{\phi}(\tau)\right\}^2 = f\{\cos\phi_1-\cos\phi(\tau)\}\label{eq_theta_1},
\end{eqnarray}
where we introduce $\phi_1\equiv\phi(1)$ and use $\dot{\phi}(1) = 0$. Equation~(\ref{eq_theta_1}) is further rewritten as
\begin{eqnarray}
\dot{\phi}(\tau) = \pm\sqrt{2f\left(\cos\phi(\tau) - \cos\phi_1\right)}\label{eq_theta_2},
\end{eqnarray}
which implies that an inflection point $\tau^*$ exists as $\phi(\tau^*) = - \phi_1$ with $0\leq\tau^*<1$. Because the strip is initially bent leftward in the present choice of symmetry for $\epsilon_x = 0$, we notice that $\dot{\phi}<0$ for $0<\tau<\tau^*$ and $\dot{\phi} > 0$ for $\tau^*<\tau<1$ would be satisfied. If the strip is bent in the opposite direction (rightward), we simply need to flip the left--right symmetry as $\varphi\to-\varphi$ and $\epsilon_x\to-\epsilon_x$.

We first show the expressions for the exact solutions in the next two subsections (Secs.~\ref{sec:inf} and \ref{sec:noinf}). Detailed derivations are presented in Sec.~\ref{app:derivation}. The exact results for the critical strain $\epsilon_x ^*$ or force-released snap onset ($\delta f_x$ and $\delta f_y$) are derived in Secs.~\ref{sec:app_cs} and \ref{sec:app_dfx}, respectively.

Before showing the exact solution, we define the elliptic integrals appearing in the exact solution~\cite{Abramowitz}. The incomplete elliptic integrals of the first and second kinds are defined as
\begin{eqnarray}
F(\phi, k) &\equiv& \int_0 ^{\phi} \frac{d\omega}{\sqrt{1-k^2\sin^2\omega}}\label{def:elliptic_F},\\
E(\phi, k) &\equiv& \int_0 ^{\sin\phi} \sqrt{\frac{1-k^2t^2}{1-t^2}}dt\label{def:elliptic_E},
\end{eqnarray}
respectively. We also introduce the amplitude function $\phi = F^{-1}(u,k) \equiv {\rm am}(u,k)$. From Eqs.~(\ref{def:elliptic_F}) and (\ref{def:elliptic_E}), we can introduce the complete elliptic integrals of the first and second kinds as $K(\kappa) \equiv F(\pi/2,\kappa)$ and $E(\kappa) \equiv E(\pi/2,\kappa)$, respectively.

\subsection{Solution with an inflection point}\label{sec:inf}

When the solution has an inflection point, i.e., the system is bistable, we obtain the following two equations for two undetermined coefficients $\varphi$ and $\phi_1$:
\begin{eqnarray}
\left(\begin{array}{c}
-\epsilon_x\\
1-\epsilon_y
\end{array}\right)
&=&
R(\varphi)
\left(\begin{array}{c}
u_x\\
u_y
\end{array}\right)\label{eq:matrix_result},\\
u_x &=&-\sqrt{\frac{2}{f}}\sqrt{\cos\varphi-\cos\phi_1}\label{eq:ux_infelct},\\
u_y &=&\frac{2}{\sqrt{f}}\left\{3E(\kappa) + E(\beta,\kappa)\right\}-1,\label{eq:uy_infelct}\\
\sqrt{f} &=& F(\beta,\kappa) + 3K(\kappa)\label{eq:sqf_result},\\
R(\varphi)&\equiv&\left(\begin{array}{cc} \cos\varphi & -\sin\varphi \\ \sin\varphi &\cos\varphi\end{array}\right).
\end{eqnarray}
Here, $F(\beta,\kappa)$ and $K(\kappa)$ are the incomplete and complete elliptic integrals of the first kind, respectively. $E(\beta,\kappa)$ and $E(\kappa)$ are the incomplete and complete elliptic integrals of the second kind, respectively [see Eqs.~(\ref{def:elliptic_F}) and (\ref{def:elliptic_E}) for the definitions]. $\kappa$ and $\beta$ are introduced as $\kappa \equiv \sin(\phi_1/2)$ and $\beta \equiv \sin^{-1}(\sin(\varphi/2)/\kappa)$, respectively.
On the basis of Eqs.~(\ref{eq:matrix_result})--(\ref{eq:sqf_result}), we can readily determine the force--strain relation as $f_y = f_y(\epsilon_x,\epsilon_y)= f\cos\varphi$ and $f_x = f_x(\epsilon_x,\epsilon_y)= f\sin\varphi$ numerically.
Note that the solution of $(\varphi,\phi_1)$ in the above equations exists if $|\epsilon_x|\leq\epsilon_x ^*$ is satisfied, where we find $\phi_1 = \pm\varphi$ at $\epsilon_x = \pm\epsilon_x ^*$. 

\subsection{Solution without inflection points}\label{sec:noinf}

For the post-snap-buckled states $|\epsilon_x|>\epsilon_x ^*$, we have a unique monostable solution with $\tau^* = 0$ in Eq.~(\ref{eq_theta_2}). If we consider the post-snap-buckled shapes with $\epsilon_x < 0$, the sign appearing in Eq.~(\ref{eq_theta_2}) is positive, and those with $\epsilon_x > 0$ are derived with a negative sign in Eq.~(\ref{eq_theta_2}). Let us show the solution for $\epsilon_x < 0$ here. Note that the results for $\epsilon_x > 0$ are derived by considering the symmetry. Writing Eq.~(\ref{eq_theta_2}) as
\begin{eqnarray}
\dot{\phi} = \sqrt{2f(\cos\phi - \cos\phi_1)}\label{eq:dphi_pos},
\end{eqnarray}
we find equations for the undetermined parameters $\varphi$ and $\phi_1$ as
\begin{eqnarray}
\left(\begin{array}{c}
-\epsilon_x\\
1-\epsilon_y
\end{array}\right)
&=&
R(\varphi)
\left(\begin{array}{c}
u_x\\
u_y
\end{array}\right)\label{eq:no_inflection1},\\
u_x &=&\sqrt{\frac{2}{f}}\sqrt{\cos\varphi - \cos\phi_1}
\label{eq:no_inflection2}\\
u_y &=&-1 -\frac{2E(\beta,\kappa) - 2E(\kappa)}{\sqrt{f}}
\label{eq:no_inflection3}\\
\sqrt{f} &=& K(\kappa) - F(\beta,\kappa)\label{eq:no_inflection4}.
\end{eqnarray}
We stress again here that Eqs.~(\ref{eq:no_inflection2})--(\ref{eq:no_inflection4}) are valid for strips without inflection points with $\epsilon_x < - \epsilon_x ^*$. Combining the above results, we can plot theoretically the exact force--strain relations in the text.  

\subsection{Detailed derivation}\label{app:derivation}
Let us derive $u_x$ and $u_y$ for solutions with an inflection point, as shown in Eqs. (\ref{eq:ux_infelct})--(\ref{eq:sqf_result}) and (\ref{eq:no_inflection2})--(\ref{eq:no_inflection4}).
We rewrite the equations for the constraints [Eq.~(\ref{x1_0_app})] as follows:
\begin{eqnarray}
\left(\begin{array}{c}
-\epsilon_x\\
1-\epsilon_y
\end{array}\right)
&=&R(\varphi)
\left(\begin{array}{c}
u_x\\
u_y
\end{array}\right),\label{eq:sin_cos_matrix_app}
\end{eqnarray}
with the integrals
\begin{eqnarray}
u_x\equiv\int_0 ^1d\tau\sin\phi(\tau),~u_y\equiv\int_0 ^1d\tau\cos\phi(\tau)\label{def:uxuy}.
\end{eqnarray}
In integrating Eq.~(\ref{def:uxuy}), we need to consider the existence of an inflection point.

\subsubsection{Solutions with an inflection point}

Here, we consider the solution with an inflection point.
Because an inflection point $\tau^*$ exists, we divide the integration region of Eq.~(\ref{def:uxuy}) into two as follows:
\begin{eqnarray}
u_x &=& u_x ^{*} + u_x ^{\dagger},\nonumber\\
u_x ^{*} &\equiv& \int_0 ^{\tau^*}d\tau\sin\phi(\tau),~u_x ^{\dagger} \equiv \int_{\tau^*} ^{1}d\tau\sin\phi(\tau),\label{def:u_x_st}\\
u_y &=& u_y ^{*} + u_y ^{\dagger},\nonumber\\
u_y ^{*} &\equiv& \int_0 ^{\tau^*}d\tau\cos\phi(\tau),~u_y ^{\dagger} \equiv \int_{\tau^*} ^{1}d\tau\cos\phi(\tau)\label{def:u_y_st}.
\end{eqnarray}
Let us integrate Eqs.~(\ref{def:u_x_st}) and (\ref{def:u_y_st}) individually using Eq.~(\ref{eq_theta_2}). First, we calculate $u_x$.
\begin{eqnarray}
u_x ^{*} &=& -\frac{1}{\sqrt{2f}}\int_{\varphi} ^{-\phi_1}\frac{d\phi}{\sqrt{\cos\phi-\cos\phi_1}}\sin\phi\nonumber\\
&=& \frac{1}{\sqrt{2f}}\int_{\cos\varphi} ^{\cos\phi_1}dt\frac{1}{\sqrt{t-\cos\phi_1}}\nonumber\\
&=&-\sqrt{\frac{2}{f}}\sqrt{\cos\varphi-\cos\phi_1},\\
u_x ^{\dagger} &=& \frac{1}{\sqrt{2f}}\int_{-\phi_1} ^{\phi_1}\frac{d\phi}{\sqrt{\cos\phi-\cos\phi_1}}\sin\phi=0.
\end{eqnarray}
Thus, we obtain
\begin{eqnarray}
u_x = -\sqrt{\frac{2}{f}}\sqrt{\cos\varphi-\cos\phi_1}\label{eq:ux_app}.
\end{eqnarray}
Next, we calculate $u_y$:
\begin{eqnarray}
u_y ^{*} &=& -\frac{1}{\sqrt{2f}}\int_{\varphi} ^{-\phi_1}\frac{d\phi}{\sqrt{\cos\phi-\cos\phi_1}}\cos\phi\nonumber\\
&=&\int_{-\phi_1} ^{\varphi}\frac{1}{\sqrt{f}\kappa}\frac{d\phi}{2}\frac{2\cos^2(\phi/2)-1}{\sqrt{1-k^2\sin^2(\phi/2)}}\nonumber\\
&=&-\frac{1}{\kappa\sqrt{f}}F(\varphi/2,1/\kappa)-\frac{1}{\kappa\sqrt{f}}F(\phi_1/2,1/\kappa)\nonumber\\
&&+\frac{1}{\sqrt{f}\kappa}\int_{-\phi_1} ^{\varphi}\frac{d\phi}{2}\frac{2\cos^2(\phi/2)}{\sqrt{1-k^2\sin^2(\phi/2)}}.
\end{eqnarray}
We remark the following two formulas for $E(\beta,k)$ and $F(\beta,k)$ with $k \equiv 1/\kappa > 1$: 
\begin{eqnarray}
\int_0 ^{a} d\omega\frac{\cos^2\omega}{\kappa\sqrt{1-k^2{\sin^2\omega}}} &=& \frac{1}{\kappa}\int_0 ^{\sin a}d(\sin\omega)\frac{\sqrt{1-\sin^2\omega}}{\sqrt{1-k^2{\sin^2\omega}}}\nonumber\\
&=& \int_0 ^{\frac{\sin a}{\kappa}}dt\sqrt{\frac{{1-\kappa^2t^2}}{1-t^2}}\nonumber\\
&=& E\left(\sin^{-1}\left(\frac{\sin a}{\kappa}\right),\kappa\right)
\label{eq:formula_E},\\
\frac{1}{\kappa}F(\omega, 1/\kappa) &=& \frac{1}{\kappa}\int_0 ^{\omega}\frac{d\tau}{\sqrt{1-\sin^2(\tau)/\kappa^2}}\nonumber\\
&=& \int_0 ^{\sin^{-1}\left(\frac{\sin\omega}{\kappa}\right)}\frac{dt}{\sqrt{1-\kappa^2\sin^2(t)}}\nonumber\\
&=& F\left(\sin^{-1}\left(\frac{\sin\omega}{\kappa}\right),\kappa\right)\label{eq:formula_F}.
\end{eqnarray}
Using Eqs.~(\ref{eq:formula_E}) and (\ref{eq:formula_F}), we find the result for $u_y ^*$:
\begin{eqnarray}
u_y ^* = \frac{2E(\beta,\kappa) + 2E(\kappa) - F(\beta,\kappa) - K(\kappa)}{\sqrt{f}}\label{eq:uy_star2}.
\end{eqnarray}
Similarly, we can push forward the calculation of $u_y ^{\dagger}$:
\begin{eqnarray}
u_y ^{\dagger} &=& \frac{1}{\sqrt{2f}}\int_{-\phi_1} ^{\phi_1}d\phi\frac{\cos\phi}{\sqrt{\cos\phi - \cos\phi_1}}\nonumber\\
&=& \sqrt{\frac{2}{{f}}}\int_{0} ^{\phi_1}d\phi\frac{\cos\phi}{\sqrt{\cos\phi - \cos\phi_1}}\nonumber\\
&=& \frac{4E(\kappa) - 2K(\kappa)}{\sqrt{f}}\label{eq:uy_dagger1}.
\end{eqnarray}
Summing Eqs.~(\ref{eq:uy_star2}) and (\ref{eq:uy_dagger1}), we find
\begin{eqnarray}
u_y = \frac{2}{\sqrt{f}}\left\{3E(\kappa) + E(\beta,\kappa)\right\} - \frac{1}{\sqrt{f}}\left\{F(\beta,\kappa) + 3K(\kappa)\right\}.\nonumber\\
\label{eq:uy_app}
\end{eqnarray}

Finally, we need to write $\sqrt{f}$ as a function of $\varphi$ and $\phi_1$. By integrating Eq.~(\ref{eq_theta_2}) in the range $0<\tau<\tau^*$, we obtain
\begin{eqnarray}
\sqrt{f}\tau^* &=& \frac{1}{\kappa} \int_{-\phi_1}^{\varphi}\frac{d\phi}{2}\frac{1}{\sqrt{1-k^2\sin^2(\phi/2)}}\nonumber\\
&=&  \frac{F(\varphi/2,1/\kappa)}{\kappa} + \frac{F(\phi_1/2,1/\kappa)}{\kappa}\nonumber\\
&=&F(\beta,\kappa) + K(\kappa)\label{eq:sqf_tstar}.
\end{eqnarray}
Furthermore, integration of Eq.~(\ref{eq_theta_2}) in the range $\tau^*<\tau<1$ reads
\begin{eqnarray}
\sqrt{f}(1-\tau^*) &=& \frac{2}{\kappa}\int_0 ^{\phi_1}\frac{d\phi}{2}\frac{1}{\sqrt{1-k^2\sin^2(\phi/2)}}\nonumber\\
&=&\frac{2F(\phi_1/2,1/\kappa)}{\kappa}\nonumber\\
&=&2K(\kappa)\label{eq:sqf_tstar_m1}.
\end{eqnarray}
The sum of Eqs.~(\ref{eq:sqf_tstar}) and (\ref{eq:sqf_tstar_m1}) is written as
\begin{eqnarray}
\sqrt{f} = F(\beta,\kappa) + 3K(\kappa)\label{eq:sqf_app}.
\end{eqnarray}
Equations (\ref{eq:sin_cos_matrix_app}), (\ref{eq:ux_app}), (\ref{eq:uy_app}), and (\ref{eq:sqf_app}) complete the exact solution for clamped--hinged elastica with an inflection point in the main text.

\subsubsection{Solutions without inflection points}
We derive the exact solution using Eq.~(\ref{eq:dphi_pos}) in a similar manner to that in the previous section. We can calculate $u_x$ and $u_y$ as follows.
\begin{eqnarray}
u_x &=& \int_0 ^1 d\tau\sin\phi\nonumber\\
&=&\frac{1}{\sqrt{2f}}\int_{\varphi} ^{\phi_1} d\phi\frac{\sin\phi}{\sqrt{\cos\phi -\cos\phi_1}}\nonumber\\
&=&\sqrt{\frac{2}{f}}\sqrt{\cos\varphi - \cos\phi_1}\label{ux_tau0},\\
u_y &=& \int_0 ^1d\tau\cos\phi\nonumber\\
&=&\frac{1}{\sqrt{2f}}\int_{\varphi} ^{\phi_1}d\phi\frac{2\cos^2(\phi/2)-1}{\sqrt{\cos\phi - \cos\phi_1}}\nonumber\\
&=&-\frac{F(\phi_1/2,1/\kappa) - F(\varphi/2,1/\kappa)}{\kappa\sqrt{f}}-\frac{2E(\beta,\kappa) - 2E(\kappa)}{\sqrt{f}}\nonumber\\
&=&\frac{F(\beta,\kappa) - K(\kappa) - 2(E(\beta,\kappa) - E(\kappa))}{\sqrt{f}}.\label{uy_tau0}
\end{eqnarray}
Integration of Eq.~(\ref{eq:dphi_pos}) from 0 to 1 reads
\begin{eqnarray}
\sqrt{f} = K(\kappa) - F(\beta,\kappa).\label{sqf_tau0}
\end{eqnarray}
Equations (\ref{eq:sin_cos_matrix_app}) and (\ref{ux_tau0})--(\ref{sqf_tau0}) complete the exact solution without inflection points, i.e., the post-snap-buckled state in the main text.

\subsection{Critical strain for snap-buckling}\label{sec:app_cs}
The critical strain for snap-buckling in the text can be derived from the exact solution as follows. The critical strain is realized when the hinged end is perpendicular to the bottom stage: $\phi_1 = \varphi$ (i.e., $\vartheta(1) = 0$). Then, 
from Eqs.~(\ref{eq:matrix_result})--(\ref{eq:sqf_result}) with $\phi_1 = \varphi$, we obtain the equations for the critical strain $\epsilon_x ^*$,
\begin{eqnarray}
-\epsilon_x ^*&=& - \left(\frac{2E(\kappa)}{K(\kappa)}-1\right)\sin\varphi
\label{eq:epx_snap},\\
1-\epsilon_y &=& \left(\frac{2E(\kappa)}{K(\kappa)}-1\right)\cos\varphi
\label{eq:epy_snap}.
\end{eqnarray}
We can numerically estimate the critical strain $\epsilon_x$, which is shown as a solid line in the main text, for any $\epsilon_y$ from Eqs.~(\ref{eq:epx_snap}) and (\ref{eq:epy_snap}). If we are interested in the small strain case $\epsilon_y\ll 1$, we can linearize Eqs.~(\ref{eq:epx_snap}) and (\ref{eq:epy_snap}) as follows. Let us expand $E(\kappa)/K(\kappa)$ in terms of small $\kappa$:
\begin{eqnarray}
\frac{2E(\kappa)}{K(\kappa)} -1&=& 1 -\kappa^2 + O(\kappa^4)\nonumber\\ 
&\simeq& 1 - \frac{\phi_1 ^2}{4} = 1 - \frac{\varphi^2}{4}\label{eq:expand_EK}.
\end{eqnarray}
Substituting Eq.~(\ref{eq:expand_EK}) into Eqs.~(\ref{eq:epx_snap}) and (\ref{eq:epy_snap}), and comparing both sides, we find the relations $\epsilon_x ^*\simeq \varphi$ and $\epsilon_y \simeq 3\varphi^2/4$. Thus, the critical strain is estimated as
\begin{eqnarray}
\epsilon_x ^*\simeq \sqrt{\frac{4\epsilon_y}{3}},\label{eq:critical_approx}
\end{eqnarray}
which is the approximate solution in the main text.

\subsection{Forces released at snap onset}\label{sec:app_dfx}
Let us derive the magnitude of the change in forces at the onset of snap.
At snap onset, the hinged angle changes discontinuously from $\vartheta(1) = 0$ to $\vartheta(1) = -2\varphi$, i.e., from $\phi_1 = \varphi$ to $\phi_1 = - \varphi$. From Eq.~(\ref{eq:sqf_app}), where we denote the amplitude of the pre- and post-snap forces as $f^+$ and $f^-$, respectively, we find the following results:
\begin{eqnarray}
\sqrt{f^+} = 4K(\kappa),~\sqrt{f^-} = 2K(\kappa).
\end{eqnarray}
Thus, the gap in the forces appearing at snap onset can be written as
\begin{eqnarray}
\delta f_y &=& 16K^2(\sin(\varphi/2))\cos\varphi\nonumber\\&& - 4K^2(\sin(-\varphi/2))\cos(-\varphi)\nonumber\\
&=& 12K^2(\kappa)\cos\varphi,\label{eq:delta_fy}\\
\delta f_x &=& 16K^2(\sin(\varphi/2))\sin\varphi \nonumber\\ &&- \left\{-4K^2(\sin(-\varphi/2))\sin(-\varphi)\right\}\nonumber\\
&=& 12K^2(\kappa)\sin\varphi,\label{eq:delta_fx}
\end{eqnarray}
where $\varphi$ is the solution of Eqs.~(\ref{eq:epx_snap}) and (\ref{eq:epy_snap}).
The final results, Eqs.~(\ref{eq:delta_fy}) and (\ref{eq:delta_fx}), are plotted as a solid line in the text. 

We derive the results for small strain, $\epsilon_y \ll 1$, from Eqs.~(\ref{eq:delta_fy}) and (\ref{eq:delta_fx}). Linearizing Eqs.~(\ref{eq:epx_snap}) and (\ref{eq:epy_snap}), we obtain $\varphi\simeq\epsilon_x ^* \simeq\sqrt{4\epsilon_y/3}$. Because the complete elliptic integral $K(\kappa)$ can be expanded in terms of $\kappa$ as $K(\kappa) = \pi/2 + \pi\kappa^2/8+\cdots$, we expand Eqs.~(\ref{eq:delta_fy}) and (\ref{eq:delta_fx}) in terms of $\epsilon_y$ as
\begin{eqnarray}
\delta f_y &=& 3\pi^2\left(1-\frac{3}{8}\varphi^2\right) + O(\varphi^4)\nonumber\\
&=& 3\pi^2\left(1-\frac{\epsilon_y}{2}\right) + O(\epsilon_y ^2),\label{eq:delta_fx_small}\\
\delta f_x &=& 3\pi^2\varphi + O(\varphi^3)\nonumber\\
&=& 2\pi^2\sqrt{3{\epsilon_y}} + O(\epsilon_y ^{3/2})\label{eq:delta_fy_small},
\end{eqnarray}
which appear in the main text.

\section{Simulation methods}\label{sec:simulation}

To investigate the validity of the experimental and theoretical results, we adopt a discrete analog of the elastic strips~\cite{discrete_model}. 
The strip is discretized into a chain of $N = 30$ particles. Because we are interested in the final shape under a given strain, we adopt the overdamped dynamics for the particles. The particles are initially aligned in a straight line from the hinged end $(0,0)$ to the clamped end $(X,Y) = (0,L)$, with sufficiently small horizontal displacement to induce initial buckling. 
The clamped-end position $(X,Y)$ is controlled with sufficiently small speed to minimize the kinetic effect. First, to realize the initial buckling of the clamped--hinged elastica, we change the height of the clamped end to give the vertical strain $\epsilon_y \equiv 1 - Y/L$. After the vertical strain is obtained, the horizontal positions of the clamped-end particles are changed to give the horizontal strain: $\epsilon_x= X/L$. After snap occurs, the reverse protocol is conducted.

\bibliography{bib_snap}

\end{document}